\documentclass[10pt,longbibliography,aps,prx,amsmath,amssymb,amsfonts,
twocolumn,superscriptaddress,footinbib,floatfix]{revtex4-2}
\usepackage{graphicx}
\usepackage{amsthm}
\newtheorem{theorem}{Theorem}
\usepackage{mathtools}
\usepackage{bm}
\usepackage{booktabs}
\usepackage{etoolbox}
\usepackage{tikz}
\usetikzlibrary{arrows.meta,positioning}
\usepackage[colorlinks=true,linkcolor=blue,citecolor=blue,urlcolor=blue]{hyperref}
\hypersetup{pdftitle={An Observable-Algebra Approach to Quantum Coherence in Anyonic Systems},
pdfauthor={Cheng-Qian Xu and D. L. Zhou}}
\apptocmd{\thebibliography}{\interlinepenalty=10000}{}{}
\newcommand{\Tr}{\operatorname{Tr}}
\newcommand{\tTr}{\widetilde{\operatorname{Tr}}}
\newcommand{\tS}{\widetilde S}

\newcommand{\M}{\mathcal M}
\newcommand{\A}{\mathcal A}
\newcommand{\ket}[1]{\lvert #1\rangle}
\newcommand{\bra}[1]{\langle #1\rvert}

\newcommand{\supp}{\operatorname{supp}}
\newcommand{\Ran}{\operatorname{Ran}}
\newcommand{\id}{\operatorname{id}}
\newcommand{\B}{\mathsf B}
\newcommand{\Cs}{C_{\rm sec}}
\newcommand{\Cf}{C_{\rm fus}}
\newcommand{\Cc}{C_{\rm conv}^{\B_A\B_B}}
\newcommand{\Cl}{C_{\rm loc}^{\B_A\B_B}}
\newcommand{\Ccc}{C_{\rm cc}^{\B_A\B_B}}
\begin{document}

\title{An Observable-Algebra Approach to Quantum Coherence in Anyonic Systems}

\author{Cheng-Qian Xu} \email{chengqianxu@hainanu.edu.cn}

\affiliation{School of Physics and Optoelectronic Engineering,
  Hainan University, Haikou 570228, China}

\author{D. L. Zhou} \email[]{zhoudl72@iphy.ac.cn}

\affiliation{Institute of Physics, Beijing National Laboratory for
  Condensed Matter Physics,\\Chinese Academy of Sciences, Beijing
  100190, China}

\affiliation{School of Physical Sciences, University of Chinese
  Academy of Sciences, Beijing 100049, China}

\date{\today}

\begin{abstract}
   Physical restrictions on operations can make quantum coherence inaccessible.
   Quantifying off-diagonal density matrix elements in a chosen basis alone
   does not reveal which information is hidden by each restriction.
   To distinguish their effects, we describe physical restrictions through
   subalgebras of accessible observables. The associated trace-preserving
   conditional expectations unify two reconstruction principles: each gives
   both the maximum-entropy state compatible with the accessible data and
   the closest state in the corresponding subalgebra in relative entropy.
   We apply this framework to anyonic systems to separate the contributions
   from local-charge superselection and non-Abelian fusion rules.
   We prove that the charge-line severing map is the conditional expectation onto
   the local-product subalgebra. It is completely positive and quantum-trace-preserving,
   and uniquely reconstructs a state in this space from all local-product expectation
   values. The hierarchy of measurement restrictions then separates total bipartite
   coherence into three nonnegative contributions: coherence between local charges,
   inaccessible fusion information, and conventional coherence. The first two decompose
   the anyonic charge entanglement and distinguish the effects of local-charge superselection
   from the remaining fusion information. Finally, we relate basis-independent measures of
   anyonic coherence to entanglement. For pure anyonic states with vacuum total charge,
   the optimized total and correlated coherence equal the total and conventional relative-entropy
   entanglement, respectively.
\end{abstract}
\maketitle

\begin{figure*}[t]
\centering
\begin{tikzpicture}[>=Stealth,font=\footnotesize,
 state/.style={draw=black!22,fill=black!1,rounded corners=2pt,
 align=center,inner sep=4pt},
 loss/.style={rounded corners=2pt,text width=3.1cm,
 minimum height=1.3cm,align=center,inner sep=6pt},
 every path/.style={line width=0.65pt}]
\definecolor{chargecolor}{HTML}{356B91}
\definecolor{fusioncolor}{HTML}{A97032}
\definecolor{basiscolor}{HTML}{367C70}
\node[state,minimum width=3.8cm,minimum height=3.25cm] (m) at (0,0) {};
\node[state,minimum width=3.4cm,minimum height=2.95cm] (bd) at (4.5,0) {};
\node[state,minimum width=3.1cm,minimum height=2.65cm] (tp) at (8.8,0) {};
\node[state,minimum width=2.8cm,minimum height=2.35cm] (di) at (12.9,0) {};
\node[font=\small] at (0,1.95) {Full state};
\node[font=\small] at (4.5,1.8) {Local-charge blocks};
\node[font=\small] at (8.8,1.65) {Local products};
\node[font=\small] at (12.9,1.5) {Local populations};
\node at (0,1.12) {$\tilde\rho$};
\node at (4.5,0.97) {$P(\tilde\rho)$};
\node at (8.8,0.82) {$E_{\rm tp}(\tilde\rho)$};
\node at (12.9,0.67) {$E_{\rm ldiag}(\tilde\rho)$};
\begin{scope}[shift={(0,-0.2)}]
\draw (-0.65,0.4) node[above] {$a$} -- (0,0.13)
 -- (0.65,0.4) node[above] {$b$};
\draw (0,0.13) -- (0,-0.13) node[midway,right] {$c$};
\draw (-0.65,-0.4) node[below] {$a'$} -- (0,-0.13)
 -- (0.65,-0.4) node[below] {$b'$};
\end{scope}
\begin{scope}[shift={(4.5,-0.2)}]
\draw[chargecolor] (-0.65,0.4) node[above] {$a$} -- (0,0.13)
 -- (0.65,0.4) node[above] {$b$};
\draw (0,0.13) -- (0,-0.13) node[midway,right] {$c$};
\draw[chargecolor] (-0.65,-0.4) node[below] {$a$} -- (0,-0.13)
 -- (0.65,-0.4) node[below] {$b$};
\end{scope}
\begin{scope}[shift={(8.8,-0.2)}]
\draw[fusioncolor] (-0.8,-0.4) node[below] {$a$} -- (-0.8,0.4) node[above] {$a$};
\draw[fusioncolor] (0.8,-0.4) node[below] {$b$} -- (0.8,0.4) node[above] {$b$};
\node[fill=white,draw=black!25,inner sep=3pt] at (-0.8,0) {$\ket{i}\bra{j}$};
\node[fill=white,draw=black!25,inner sep=3pt] at (0.8,0) {$\ket{k}\bra{l}$};
\end{scope}
\begin{scope}[shift={(12.9,-0.2)}]
\draw[basiscolor] (-0.8,-0.4) node[below] {$a$} -- (-0.8,0.4) node[above] {$a$};
\draw[basiscolor] (0.8,-0.4) node[below] {$b$} -- (0.8,0.4) node[above] {$b$};
\node[fill=white,draw=basiscolor!45,inner sep=3pt] at (-0.8,0) {$\ket{i}\bra{i}$};
\node[fill=white,draw=basiscolor!45,inner sep=3pt] at (0.8,0) {$\ket{k}\bra{k}$};
\end{scope}

\draw[->,chargecolor] (m.east) -- node[above=3pt] {$P$} (bd.west);
\draw[->,fusioncolor] (bd.east) -- node[above=3pt] {$E_{\rm tp}$} (tp.west);
\draw[->,basiscolor] (tp.east) -- node[above=3pt] {$E_{\rm ldiag}$} (di.west);
\draw[draw=black!28,fill=black!2,rounded corners=4pt]
 (0.4,-2.2) rectangle (8.6,-4.25);
\node[align=center,text=black!80] at (4.5,-3.95)
 {$S_{\rm ACE}$: charge entanglement};
\node[loss,fill=chargecolor!9,draw=chargecolor!65] (sec) at (2.35,-3.05)
 {{\small\textcolor{chargecolor}{$C_{\rm sec}$}}\\Charge-sector coherence};
\node[font=\Large,text=black!55] at (4.5,-3.05) {$+$};
\node[loss,fill=fusioncolor!9,draw=fusioncolor!65] (fus) at (6.7,-3.05)
 {{\small\textcolor{fusioncolor}{$C_{\rm fus}$}}\\Fusion information};
\node[font=\Large,text=black!55] at (8.85,-3.05) {$+$};
\node[loss,fill=basiscolor!9,draw=basiscolor!65] (conv) at (10.9,-3.05)
 {{\small\textcolor{basiscolor}{$C_{\rm conv}^{\B_A\B_B}$}}\\Conventional coherence};
\draw[->,chargecolor!70] (2.35,-0.12) -- (2.35,-2.08);
\draw[->,fusioncolor!70] (6.7,-0.12) -- (6.7,-2.08);
\draw[->,basiscolor!70] (10.9,-0.12) -- (conv.north);
\node[font=\Large,text=black!55] at (12.95,-3.05) {$=$};
\node[loss,text width=1.9cm,draw=black!50,fill=black!5] at (14.35,-3.05)
 {{\small $C^{\B_A\B_B}$}\\Total\\\mbox{coherence}};
\end{tikzpicture}
\caption{Reconstruction and coherence decomposition. From left to right, successive reconstructions
remove charge-sector coherence, unresolved fusion information, and conventional coherence.
Smaller frames indicate less retained information, with no quantitative size scale.
Each contribution below is an entropy increase; the first two form $S_{\rm ACE}$, and all three
sum to the total bipartite coherence. Diagrams represent operator components with coefficients
and sums omitted. Internal fusion trees and multiplicity labels are suppressed; $i,j$ and
$k,l$ label local basis states. Disconnected lines denote local operator products, not necessarily
a product state.}
\label{fig:hierarchy}
\end{figure*}
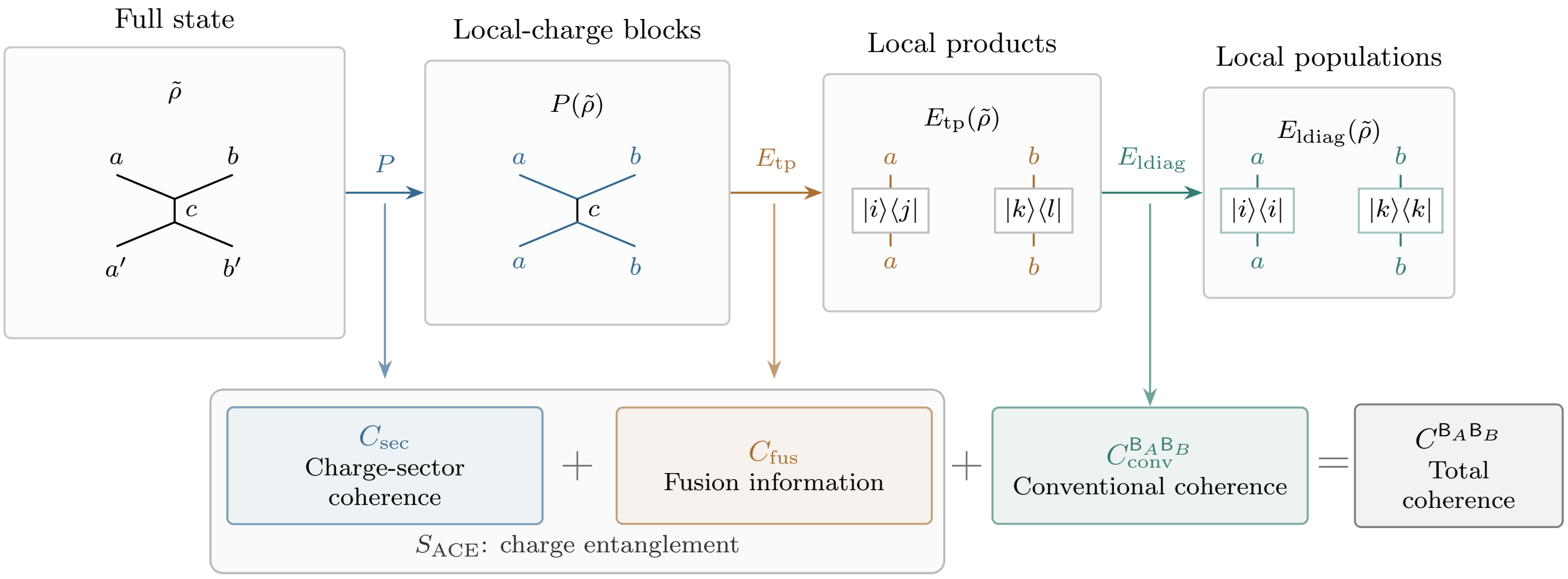

\section{Introduction}
\label{sec:introduction}

Quantum coherence~\cite{Baumgratz2014,Levi_2014,RevModPhys.89.041003,Lecamwasam2024,Mani2024,Ahnefeld2026}
is usually described by the off-diagonal elements of a density matrix in a chosen basis.
In real constrained systems, however, not all of this coherence is accessible:
physical restrictions limit the state transformations and measurements that can be performed.
Superselection rules (SSRs)~\cite{BartlettWiseman2003,RevModPhys.79.555}, for example, forbid operations that connect
different sectors. Coherence between these sectors can be inaccessible to
observers whose operations obey the SSR.
Off-diagonal elements alone therefore do not tell us which coherence is accessible
under the allowed operations. Distinguishing the information hidden by different restrictions
requires considering the allowed operations rather than the state alone.

Anyons are excitations in two-dimensional systems~\cite{PhysRevLett.49.957,PhysRevLett.48.1144}.
Quantum information can be encoded in their fusion degrees of
freedom~\cite{kitaev2006anyons,pachos_2012,Nayak2008}.
Accessing this information through local measurements is subject to two restrictions.
First, anyonic SSRs~\cite{PhysRevA.69.052326} require observables to preserve each subsystem's
total topological charge, making coherence between different local charges inaccessible.
Second, non-Abelian fusion rules allow two given charges to combine in different ways.
The resulting fusion information is not determined by the states of the two subsystems separately
and can remain inaccessible even when all correlations between local measurements are
recorded~\cite{XuZhou2023}. These restrictions call for a distinction between
coherence across local-charge sectors and the information associated with their fusion.

Previous studies quantified information inaccessible to local observers through
anyonic charge correlations. A probe of these correlations was defined through a charge-line
severing map~\cite{Bonderson2017}. For full-rank states, a maximum-entropy approach identified
this quantity with the information that cannot be recovered from local measurement
data~\cite{XuZhou2023}. Relative-entropy entanglement measures further separated
conventional entanglement from anyonic charge entanglement~\cite{4nd3-k6rw}.
These results quantify the information inaccessible to local observers, but leave open how to
distinguish the contributions from local-charge superselection rules and non-Abelian fusion rules.
Our aim is to separate these two contributions and clarify how they combine with
conventional coherence to give the total bipartite coherence.

We approach this problem through the algebra of accessible
observables~\cite{Furuya2020,Bertoni2020,Casini2021,Zanardi2004,Balachandran2013},
rather than through the matrix elements of the state alone. The reconstruction framework depends
on the accessible observables and does not require anyonic structure. We apply it here
to distinguish the information hidden by superselection and fusion constraints.
When these observables form a subalgebra, their trace-preserving conditional expectation
defines a unique linear reconstruction that lies in the subalgebra and preserves all
accessible expectation values. This reconstruction unifies two prescriptions:
it is the maximum-entropy state compatible with the measurement data and the closest state
in the subalgebra in relative entropy. For anyonic systems, we prove that the charge-line severing
map is the conditional expectation onto the local-product subalgebra. It uniquely reconstructs
a state in the local-product subalgebra while preserving all local-product expectation values,
and it is completely positive and quantum-trace-preserving.

Figure~\ref{fig:hierarchy} summarizes the resulting decomposition. A local-charge
projection first removes coherence between different local charge sectors. Reconstruction
from local-product data then removes the remaining inaccessible fusion information.
Finally, retaining only populations in fixed local bases removes conventional
coherence. The corresponding subalgebras give an additive decomposition into three
nonnegative contributions for both pure and mixed states. In particular, the first
two contributions separate the entropy of anyonic charge entanglement according to
the two restrictions above. We further divide conventional coherence into local
and correlated parts. For pure states with vacuum total charge, optimized total
and correlated coherence recover the total and conventional relative-entropy
entanglement, respectively.

\section{Reconstructing a state from restricted measurements}
\label{sec:framework}

A density matrix predicts the outcome of every measurement allowed by quantum mechanics.
An observer with access to only some of those measurements generally cannot determine the full state.
For example, measuring a spin along $z$ determines its populations in the $\{\ket0,\ket1\}$ basis,
but gives no information about the coherence between them. The question is then not only how much
coherence the state contains, but which information is missing from the observer's description.

We approach this question in two ways. First, among all states that reproduce the same accessible
measurement data, the maximum-entropy principle selects the state that adds the least information
beyond those data~\cite{Jaynes1957,PhysRevLett.89.207901,PhysRevLett.101.180505,Teo2011,Gupta2021}.
Second, we ask how well the original state can be approximated by a state described entirely
in terms of the accessible observables, using quantum relative entropy to quantify the
difference~\cite{930911,PhysRevA.80.022113,Gao2020}. These are two distinct optimization problems. Below
we explain their physical meaning and the condition under which they give the same reconstructed state.

We begin with ordinary density matrices, $\rho \ge 0$ and $\Tr \rho = 1$.
Entropy and relative entropy are $S(\rho) = - \Tr(\rho \log_2 \rho)$
and $S(\rho \Vert \sigma) = \Tr[\rho(\log_2 \rho - \log_2 \sigma)]$.
We set $S(\rho\Vert\sigma)=+\infty$ if $\supp\rho\nsubseteq\supp\sigma$.
The anyonic version uses the quantum trace $\tTr$ introduced in Sec.~\ref{sec:anyonic-algebras}.

\subsection{Maximum-entropy inference from incomplete data}
\label{subsec:maxent}

Suppose an observer knows the expectation values of the accessible observables $O$.
Denote their operator space by $\A$. The compatible states satisfy
\begin{equation}
   \mathcal{K}_{\A}(\rho) = \left\{ \sigma \ge 0 \ \middle|\
   \begin{aligned}
      &\Tr\sigma = 1,\\
      &\Tr(\sigma O) = \Tr(\rho O),\quad \forall O \in \A
   \end{aligned}\right\}.
   \label{eq:compatible-states}
\end{equation}
Every state in this set gives the same accessible measurement statistics, although
they may differ in information that the observer cannot measure. Maximum-entropy inference selects
\begin{equation}
   \rho_{\A}^{\max} = \underset{\sigma \in \mathcal{K}_{\A}(\rho)}{\arg\max}\ S(\sigma).
   \label{eq:maxent-definition}
\end{equation}
This prescription does not claim that the actual state has maximum entropy.
It assigns an effective state using only the accessible data,
without retaining additional coherence or correlations that those data do not determine.

For example, if only the populations $p$ and $1-p$ of a qubit are known,
its compatible density matrices have the form
\begin{equation}
   \sigma = \begin{pmatrix}
               p & z \\
               z^* & 1-p
            \end{pmatrix}, \qquad |z|^2 \le p(1-p).
   \label{eq:qubit-compatible}
\end{equation}
The measured populations leave $z$ undetermined. Entropy is largest for $z=0$,
so the inferred state is $\rho_{\A}^{\max} = {\rm diag}(p,1-p)$, with entropy $H_2(p)$,
which is the binary Shannon entropy. Its excess entropy over the actual state quantifies the
information missing from population measurements.

\subsection{The closest state in the accessible description}
\label{subsec:relative-projection}

The second approach starts with the full state $\rho$ and compares it with states
represented within the accessible operator space
\begin{equation}
   \mathcal{S}(\A) = \{\sigma \in \A: \sigma \ge 0, \ \Tr \sigma = 1\}.
\end{equation}
The relative-entropy approximation is
\begin{equation}
   \rho_{\A}^{\rm proj} = \underset{\sigma \in \mathcal{S}(\A)}{\arg\min}\ S(\rho\Vert\sigma).
   \label{eq:relative-projection}
\end{equation}
Here the candidate states are restricted by their form, rather than by requiring them
to reproduce the same measurement data. In the qubit example above, these candidates
are all diagonal states $\sigma_q = {\rm diag}(q, 1-q)$, with arbitrary $q$.
Thus, $S(\rho \Vert \sigma_q) = -S(\rho) - p\log_2q - (1-p)\log_2(1-q)$,
which is minimized at $q=p$. Thus the closest diagonal state has the same populations as $\rho$,
even though that requirement was not imposed in Eq.~\eqref{eq:relative-projection}.

The phrase ``closest state'' refers throughout to this minimization of $S(\rho\Vert\sigma)$.
Relative entropy is not a metric: it is asymmetric and does not obey the usual triangle inequality.
What matters here is that it is nonnegative and vanishes only when the two states coincide.

\subsection{State reconstruction from accessible observables}
\label{subsec:reconstruction-map}

The qubit example can be extended to a broad class of measurement restrictions. We assume
that $\A$ contains the identity and is closed under linear combinations, products, and Hermitian
conjugation: it is an observable subalgebra. This condition is important. It lets us construct
a reconstruction map $E_{\A}$ that retains all expectation values available to the observer,
\begin{equation}
   \Tr[E_{\A}(\rho) O] = \Tr(\rho O),\qquad \forall O \in \A,
   \label{eq:pairing}
\end{equation}
while producing a state $E_{\A}(\rho)$ within $\A$. The map is the trace-preserving
\emph{conditional expectation} onto $\A$~\cite{Umegaki1954,GaoRouze2022}.
It is linear, completely positive, and leaves states already in $\A$ unchanged. Requiring
the output to lie in $\A$ and preserve its expectation values determines the map uniquely.
Appendix~\ref{app:expectation} gives the construction.

The state $E_{\A}(\rho)$ minimizes the relative entropy over states in $\A$~\cite{Gao2020}.
As shown below, this relation also implies that $E_{\A}(\rho)$ has the largest entropy
among all states consistent with the accessible measurement data.

\begin{theorem}[State reconstruction]
\label{thm:reconstruction}
Let $\A$ be a finite-dimensional observable subalgebra containing the identity, and let $E_{\A}$
be its trace-preserving conditional expectation. For any density matrix $\rho$,
the maximum-entropy state $\rho^{\rm max}_{\A}$ compatible with its accessible measurement data
and the state $\rho_{\A}^{\rm proj}$ in $\mathcal S(\A)$ that minimizes relative entropy
from $\rho$ are both unique and satisfy
\begin{equation}
   \rho_{\A}^{\max} = \rho_{\A}^{\rm proj} = E_{\A}(\rho).
   \label{eq:two-projections}
\end{equation}
\end{theorem}
Moreover, the relative entropy between $\rho$ and its reconstructed state $E_{\A}(\rho)$
equals the increase in entropy under reconstruction,
\begin{equation}
   S[\rho\Vert E_{\A}(\rho)] = S[E_{\A}(\rho)] - S(\rho).
   \label{eq:entropy-increase}
\end{equation}
A detailed proof of both statements, including the case of states with zero eigenvalues,
is given in Appendix~\ref{app:expectation}.

We therefore use
\begin{equation}
   C_{\A}(\rho) := S[E_{\A}(\rho)] - S(\rho) = S[\rho\Vert E_{\A}(\rho)]
   \label{eq:algebraic-coherence}
\end{equation}
as the information lost in this restricted description~\cite{Liu2017}. When $\A$ consists of
operators diagonal in a chosen basis, $E_{\A}$ is dephasing and this quantity is the usual
relative-entropy coherence~\cite{Baumgratz2014}. For the qubit state $\ket+ = (\ket0 + \ket1)/\sqrt2$,
population measurements reconstruct $I/2$ and lose one bit of information. More general restrictions
can also erase correlations. For instance, if only subsystem $A$ can be measured,
the reconstructed state of a bipartite system is
\begin{equation}
   E_{\A}(\rho_{AB}) = \rho_A \otimes \frac{I_B}{\dim\mathcal{H}_B},
   \qquad \A = \mathcal{B}(\mathcal{H}_A) \otimes I_B.
\end{equation}
Here $\mathcal{B}(\mathcal{H}_A)$ denotes the algebra of all linear operators
on the Hilbert space of subsystem $A$, and $I_B$ is the identity operator on $B$.
The reconstruction preserves everything measurable on $A$ while discarding information about $B$
and its correlations with $A$. Accordingly, we use ``information loss''
when the restriction is more general than basis dephasing. Appendix~\ref{app:ordinary}
gives familiar examples from ordinary coherence and discord.

\subsection{Hierarchy of information loss}
\label{subsec:stages}

Now compare two observers. One has access to a larger set of observables $\mathcal{B}$,
while the other can measure only a subset $\A \subseteq \mathcal{B}$. Both sets are assumed
to be subalgebras. Reconstructing from the larger set does not change any data needed
for reconstruction from the smaller set. The corresponding maps therefore satisfy
\begin{equation}
   E_{\A} E_{\mathcal{B}} = E_{\mathcal{B}} E_{\A} = E_{\A}.
   \label{eq:tower}
\end{equation}
The total information loss separates into two nonnegative parts,
\begin{align}
   S[\rho \Vert E_{\A}(\rho)] = S[\rho \Vert E_{\mathcal{B}}(\rho)]
      + S[E_{\mathcal{B}}(\rho) \Vert E_{\A}(\rho)].
   \label{eq:staged-loss}
\end{align}
The first term measures the information lost when the description is restricted to $\mathcal{B}$.
The second term measures the further loss when the description is restricted from $\mathcal{B}$
to $\A$. These reconstruction results apply to finite-dimensional quantum systems
whenever the accessible observables form subalgebras containing the identity,
with all reconstructions preserving the same trace. The information lost at each
step depends on the physical meaning of the corresponding measurement restriction.

\section{Anyonic constraints on local measurements}
\label{sec:anyonic-algebras}

We now apply the framework developed above to anyonic systems. Superselection rules
and fusion rules restrict the information accessible to local measurements. As a result,
products of local observables generally do not span the full physical operator space.
In this section, we introduce the anyonic state space and identify four observable algebras:
the full physical algebra, the algebra preserving each local charge, the algebra generated
by local products, and its diagonal subalgebra in chosen local bases. Their inclusion relations
provide the basis for the coherence decomposition in Sec.~\ref{sec:decomposition}.

\subsection{Fibonacci anyons, fusion spaces, and anyonic states}
\label{subsec:fibonacci-basis}

We use Fibonacci anyons as a concrete example to introduce fusion spaces,
fusion-tree bases, and the operator and state conventions needed below.
More detailed introductions can be found in
Refs.~\cite{kitaev2006anyons,trebst2008short,pachos_2012,bkn5-drgw}.

The Fibonacci model has two topological charges, $\{1,\tau\}$. Here $1$ denotes the vacuum
and $\tau$ the Fibonacci anyon. Their fusion rules are
\begin{align}
   1 \times \tau & = \tau \times 1 = \tau, \nonumber\\
   \tau \times \tau & = 1 + \tau.
\end{align}
Thus, two $\tau$ anyons can have total charge $1$ or $\tau$.
Their quantum dimensions are $d_1 = 1$ and $d_\tau = (1+\sqrt5)/2$.

When charges $a$ and $b$ combine to give total charge $c$, the possible fusion channels
span a space denoted by $V_{ab}^{c}$. Its dimension is the fusion multiplicity $N_{ab}^{c}$,
and a basis vector is written as $\ket{a,b;c,\mu}$, with $\mu=1,\ldots,N_{ab}^{c}$.
The Fibonacci model is multiplicity-free: every $N_{ab}^{c}$ is either zero or one,
so the label $\mu$ can be omitted.
The two allowed fusion channels of two $\tau$ anyons are represented by
\begin{align}
    & \ket{\tau, \tau; 1} = \left(\frac{1}{d_\tau}\right)^{\frac{1}{2}}
    

\begin{tikzpicture}
        \draw (-0.5,0.25) -- (0,-0.25) -- (0.5,0.25);
    \end{tikzpicture}, \nonumber \\
    & \ket{\tau, \tau; \tau} = \left(\frac{1}{d_\tau}\right)^{\frac{1}{4}}
    

\begin{tikzpicture}[baseline]
        \draw (-0.5,0.5) -- (0,0) -- (0.5,0.5);
        \draw (0,0) -- (0,-0.5);
    \end{tikzpicture}. \nonumber
\end{align}
We use the standard graphical notation: world lines are oriented upward (arrows are omitted),
a solid line represents $\tau$, and vacuum lines are not drawn. The superselection rule
forbids physical operations from changing the total charge. When the total charge is not fixed,
the two-anyon Hilbert space is
therefore $\mathcal{H}_{\tau\tau} = V_{\tau\tau}^{1} \oplus V_{\tau\tau}^{\tau}$.

For more than two anyons, the order of fusion is not unique. Each fusion order is represented
by a fusion tree and defines a basis of the fusion space. Consider an ordered collection
with charges $\vec a=(a_1,a_2,\ldots,a_n)$ and fixed total charge $c$. We denote its fusion space
by $\mathcal{H}_{\vec a}^{c} \equiv V_{a_1 a_2 \cdots a_n}^{c}$ and, unless stated otherwise,
use the left-associated fusion tree. A basis vector is
\begin{equation}
   \ket{\vec{a}; \vec{x}; c} \equiv \ket{a_1, a_2; x_1} \otimes \ket{x_1, a_3; x_2}
      \otimes \cdots \otimes \ket{x_{n-2}, a_n; c}. \nonumber
\end{equation}
Here $\vec{x} = (x_1, x_2, \ldots, x_{n-2})$ lists the intermediate charges.
For models with fusion multiplicities, each vertex also carries a multiplicity label.
Other fusion trees give other bases of the same fusion space.

Given two anyonic fusion spaces $\mathcal{H}_A$ and $\mathcal{H}_B$, the space of
anyonic operators from $\mathcal{H}_A$ to $\mathcal{H}_B$ can be represented
as $\mathcal{B}(\mathcal{H}_A \to \mathcal{H}_B) \simeq \mathcal{H}_B \otimes \mathcal{H}_A^{*}$.
Within a fixed total-charge sector, anyonic operators act as ordinary matrices
on the corresponding fusion space. When several total-charge sectors are allowed,
physical operators are block diagonal in the total charge. We use a tilde, $\tilde{O}$,
to denote an anyonic operator.

The trace operation appropriate for anyonic operators $\tilde{O}$ is the quantum trace,
denoted by $\tTr$, defined as $\tTr(\tilde{O}) = \sum_c d_c \Tr(\tilde{O}_c)$,
where $\tilde{O}_c$ denotes the component of $\tilde{O}$ in charge sector $c$,
and $\Tr$ is the ordinary matrix trace within that sector.
An anyonic state is a positive anyonic operator normalized by the quantum trace.
The corresponding anyonic entropy and anyonic relative entropy are defined
as $\tS(\tilde\rho) = -\tTr(\tilde{\rho} \log_2 \tilde{\rho})$
and $\tS(\tilde{\rho} \|\tilde {\sigma})
   = \tTr(\tilde{\rho} \log_2 \tilde{\rho}) - \tTr(\tilde{\rho} \log_2 \tilde{\sigma})$,
with the usual support convention. These are the usual definitions with the trace appropriate
to the anyonic state space~\cite{Bonderson2017}. The reconstruction results of
Sec.~\ref{sec:framework} also hold for this trace, which is explained
in Appendix~\ref{app:expectation}.

\subsection{Information accessible to local measurements}
\label{subsec:product}

For an ordinary bipartite system, measuring a complete set of observables on each side,
and recording their joint correlations, is sufficient to reconstruct the density matrix.
This follows from
\begin{equation}
   \mathcal{B}(\mathcal{H}_A \otimes \mathcal{H}_B) = \mathcal{B}(\mathcal{H}_A) \otimes \mathcal{B}(\mathcal{H}_B).
\end{equation}
The analogous measurements can not determine an anyonic state in general: local observables preserve
each subsystem's total charge, and the way the two charges fuse is not an internal degree of
freedom of either subsystem.

Consider a bipartite anyonic system with subsystems $A$ and $B$. Let $a$ and $b$ be their
total charges, and let $c$ be the total charge of $AB$. The state of $AB$ specifies
the internal fusion states in $A$ and $B$, as well as how $a$ and $b$ fuse. Accordingly,
\begin{equation}
   \mathcal{H}_{AB}^{c} = \bigoplus_{a,b} \mathcal{H}_A^a \otimes \mathcal{H}_B^b \otimes V_{ab}^{c}.
   \label{eq:bipartite-space}
\end{equation}
Here $\mathcal{H}_A^a$ and $\mathcal{H}_B^b$ contain the internal states, and $V_{ab}^{c}$
describes the fusion space. Only charge pairs allowed by the fusion rules are included.

The global superselection rule requires physical operators to preserve the total charge $c$.
We denote the algebra of these operators by
\begin{equation}
   \M = \bigoplus_c \mathcal{B}(\mathcal{H}_{AB}^{c}).
   \label{eq:ambient}
\end{equation}
Within the same $c$ sector, coherence between different local charge pairs $(a,b)$ is allowed.
For example, a global vacuum state can superpose local charges $(1,1)$ and $(\tau,\tau)$.
Both pairs can fuse to the vacuum. However, such coherence is invisible to local observables
that preserve the charge in each subsystem separately.

We assume access to all charge-preserving observables within each subsystem, including their
joint measurement statistics. Their products generate $\A_{\rm tp}\simeq\A_A\otimes\A_B$, where
\begin{equation}
   \A_A = \bigoplus_a \mathcal{B}(\mathcal{H}_A^a),\qquad \A_B = \bigoplus_b \mathcal{B}(\mathcal{H}_B^b).
\end{equation}
On a block with charges $(a,b,c)$, an observable in $\A_{\rm tp}$ acts
as $O_{ab} \otimes I_{V_{ab}^{c}}$. Here $O_{ab}$ acts on the internal states of $A$ and $B$,
while $I_{V_{ab}^{c}}$ is the identity on their fusion space. Thus these measurements cannot
detect coherence between different local charge pairs. They also can not determine the state
in $V_{ab}^{c}$ or distinguish the possible total charges $c$ at fixed $(a,b)$. This remains
true even when the observers record all correlations between their local measurements.

To separate these two restrictions, first consider a measurement of the total charge in each
subsystem with its outcome discarded. If $\Pi_{ab}$ projects onto local charges $(a,b)$,
the state after this measurement is
\begin{equation}
   P(\tilde\rho) = \sum_{a,b} \Pi_{ab} \tilde\rho \Pi_{ab}.
   \label{eq:P}
\end{equation}
This removes coherence between different local charge pairs and preserves their probabilities.
It leaves all matrix elements within each fixed $(a,b)$ block unchanged,
including the information in fusion space.

The operators left unchanged by $P$ form the subalgebra
\begin{equation}
   \A_{\rm bd} = \Ran P = \bigoplus_c \bigoplus_{a,b}
      \mathcal{B}( \mathcal{H}_A^a \otimes \mathcal{H}_B^b \otimes V_{ab}^{c}).
 \label{eq:bd-algebra}
\end{equation}
Here we denote this block-diagonal algebra by $\A_{\rm bd}$, and $\Ran P$ denotes the set of
outputs of $P$. Sums, products, and Hermitian conjugates of operators that preserve $(a,b)$
still preserve $(a,b)$. The map $P$ preserves the quantum trace and every expectation value
in $\A_{\rm bd}$. It is thus the conditional expectation onto $\A_{\rm bd}$,
so the reconstruction principle of Sec.~\ref{sec:framework} applies.
Appendix~\ref{app:expectation} gives the proof.

The algebra $\A_{\rm bd}$ includes joint observables that distinguish how $a$ and $b$ fuse.
Such observables generally cannot be formed from local products, so
\begin{equation}
   \A_{\rm tp} \subseteq \A_{\rm bd} \subseteq \M.
   \label{eq:tp-bd}
\end{equation}

To retain only the information accessible to local products, let $E_{\rm tp}(\tilde\rho)$
be the state reconstructed from all local-product expectation values. The reconstruction uses
only local-product data, and a known global charge is not imposed as an additional constraint.
Its defining property is
\begin{equation}
   \tTr[E_{\rm tp}(\tilde\rho) O] = \tTr(\tilde\rho O), \qquad \forall O \in \A_{\rm tp},
   \label{eq:local-data}
\end{equation}
with $E_{\rm tp}(\tilde\rho) \in \A_{\rm tp}$. By Theorem~\ref{thm:reconstruction},
it is both the maximum-entropy state consistent with these data and the closest state
in $\A_{\rm tp}$ based on relative entropy.

For example, consider two Fibonacci anyons, with one $\tau$ in each region. The local charges
are fixed, so $P$ leaves the state unchanged. Nevertheless, the pair can have total charge $1$
or $\tau$. Neither local observer can determine which from measurements on a single $\tau$.
All mixtures of these two fusion outcomes give the same local data. Reconstruction from
those data must therefore assign the same state to all of them. We calculate the resulting
information loss in Sec.~\ref{subsec:two-tau}.

More generally, for known local charges $(a,b)$, the reconstructed probability of
fusion outcome $c$ is
\begin{equation}
   p_{ab}(c) = \frac{N_{ab}^{c}d_c}{d_a d_b}.
   \label{eq:fusion-reference-distribution}
\end{equation}
This distribution follows from maximizing the anyonic entropy. It is normalized
because $\sum_c N_{ab}^{c}d_c=d_a d_b$. For fixed $(a,b)$, local-product measurements determine
an averaged internal state of $A$ and $B$. They cannot determine the fusion outcome $c$,
the state within $V_{ab}^{c}$, or correlations between the fusion and internal degrees of freedom.
The reconstruction $E_{\rm tp}$ therefore uses the same reconstructed internal state
for every allowed $c$, assigns $c$ the probability $p_{ab}(c)$, and assigns the maximally
mixed state to $V_{ab}^{c}$. These changes remove the fusion information retained by $P$.

Since local products never detect coherence between different $(a,b)$, applying $P$ first
does not alter their data. Consequently,
\begin{equation}
   E_{\rm tp} P = P E_{\rm tp} = E_{\rm tp}.
   \label{eq:P-E}
\end{equation}

The charge-line severing map $\Omega^{A:B}$, defined using an $\Omega$-loop, was introduced
to quantify anyonic charge entanglement~\cite{Bonderson2017,XuZhou2023,4nd3-k6rw}. It removes
nontrivial charge lines connecting $A$ and $B$. In the multiplicity-free case, its action is
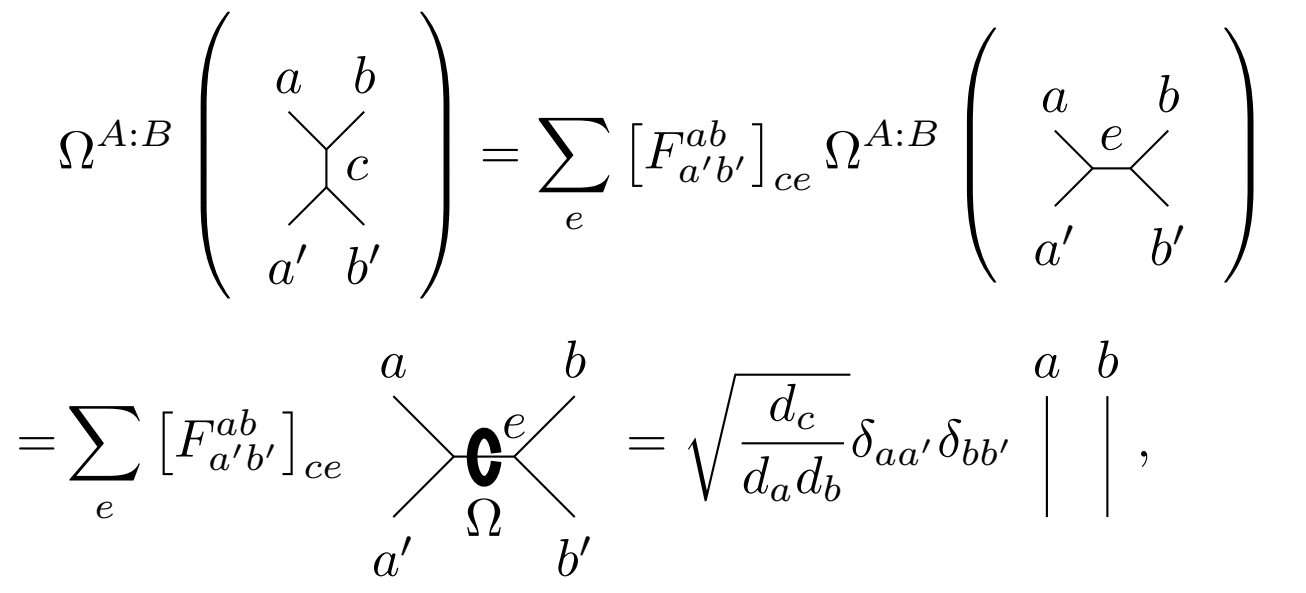
\begin{align}
    & \Omega^{A:B} \left(~ \begin{tikzpicture}[baseline, scale=0.5]
        \draw (-0.5,0.75) node[above]{$a$} -- (0,0.25) -- (0.5,0.75) node[above]{$b$};
        \draw (-0.5,-0.75) node[below]{$a'$} -- (0,-0.25) -- (0.5,-0.75) node[below]{$b'$};
        \draw (0,0.25) -- (0,-0.25) node[pos=0.5, right]{$c$};
    \end{tikzpicture}~ \right)
    = \sum_e \left[ F^{ab}_{a' b'} \right]_{ce} \Omega^{A:B} \left(~ \begin{tikzpicture}[baseline, scale=0.5]
        \draw (-0.75,0.5) node[above]{$a$} -- (-0.25,0) -- (-0.75,-0.5) node[below]{$a'$};
        \draw (0.75,0.5) node[above]{$b$} -- (0.25,0) -- (0.75,-0.5) node[below]{$b'$};
        \draw (-0.25,0) -- (0.25,0) node[pos=0.5, above]{$e$};
    \end{tikzpicture}~ \right) \nonumber \\
    = & \sum_e \left[ F^{ab}_{a' b'} \right]_{ce}
    \begin{tikzpicture}[baseline, scale=0.8]
        \draw (-0.75,0.5) node[above]{$a$} -- (-0.25,0) -- (-0.75,-0.5) node[below]{$a'$};
        \draw (0.75,0.5) node[above]{$b$} -- (0.25,0) -- (0.75,-0.5) node[below]{$b'$};
        \draw (-0.25,0) -- (0.25,0) node[above]{$e$};
        \draw[line width=2pt] (0.1,0.03) arc(10:350:0.1 and 0.2);
        \draw (0,-0.2) node[below]{$\Omega$};
    \end{tikzpicture} = \sqrt{\frac{d_c}{d_a d_b}} \delta_{aa'} \delta_{bb'}
    \begin{tikzpicture}[baseline, scale=0.8]
        \draw (-0.25,-0.5) -- (-0.25,0.5) node[above]{$a$};
        \draw (0.25,-0.5) -- (0.25,0.5) node[above]{$b$};
    \end{tikzpicture}, \nonumber
\end{align}
where the first step uses an $F$ move, a basis transformation between different fusion orders.
The internal fusion trees in subsystems $A$ and $B$ are suppressed.
The $\Omega$-loop projects the connecting charge onto the vacuum.

\begin{theorem}[Reconstruction by charge-line severing]
\label{thm:severing}
For the bipartite anyonic system considered here, the charge-line severing map is the
quantum-trace-preserving conditional expectation onto the local-product algebra $\A_{\rm tp}$:
\begin{equation}
   \Omega^{A:B} = E_{\rm tp}.
   \label{eq:severing-identification}
\end{equation}
Consequently, for any state $\tilde\rho$, $\Omega^{A:B}(\tilde\rho)$ is the unique
maximum-entropy state consistent with all local-product expectation values. It is also
the unique state in $\A_{\rm tp}$ that minimizes the relative entropy from $\tilde\rho$.
\end{theorem}

The identification is proved in Appendix~\ref{app:expectation}; the two reconstruction
statements then follow from Theorem~\ref{thm:reconstruction} with the quantum trace.


Finally, consider restricting measurements to a fixed local basis. Choose an orthonormal
basis $\B_A^a$ in each $\mathcal{H}_A^a$ and $\B_B^b$ in each $\mathcal{H}_B^b$.
Fusion-tree bases are one such choice. Each basis state has a definite local charge.
The observers now record only the joint probabilities of these basis states.

These data define the local diagonal subalgebra
\begin{equation}
   \A_{\rm ldiag}^{\B_A\B_B} = \A_{A,\rm diag}^{\B_A} \otimes \A_{B,\rm diag}^{\B_B} \subseteq \A_{\rm tp},
   \label{eq:ldiag}
\end{equation}
where $\A_{A,\rm diag}^{\B_A}$ consists of operators diagonal in the chosen bases of $A$,
and similarly for $B$. The tensor product acts in the global fusion space as described above.
Its reconstruction map $E_{\rm ldiag}$ first retains the local-product data and then dephases
the local internal states in the selected bases. We suppress the basis labels on this map
when the bases are fixed.

The observable algebras satisfy
\begin{equation}
   \A_{\rm ldiag}^{\B_A\B_B} \subseteq \A_{\rm tp} \subseteq \A_{\rm bd} \subseteq \M.
   \label{eq:hierarchy}
\end{equation}
These inclusions describe successive restrictions on the measurements. The full algebra $\M$
respects global anyonic superselection rules. The algebra $\A_{\rm bd}$ also preserves each
local charge, while $\A_{\rm tp}$ contains only information accessible to local products.
Finally, $\A_{\rm ldiag}^{\B_A\B_B}$ retains only populations in the chosen local bases.
In the next section we quantify the information lost at these steps.

\section{Decomposition of bipartite anyonic coherence}
\label{sec:decomposition}

The observable restrictions identified in Sec.~\ref{sec:anyonic-algebras} allow us to
distinguish different contributions to anyonic coherence. We begin with coherence
in a global fusion basis. We then define bipartite coherence from local measurement data
and separate it into three contributions: coherence between local charges, fusion information
inaccessible to local measurements, and conventional coherence within local charge sectors.

\subsection{Fusion-basis coherence and the local-measurement decomposition}
\label{subsec:global-coherence}

We first consider coherence in a fixed global fusion basis. As in an ordinary quantum system,
the diagonal entries specify populations, whereas the off-diagonal entries describe coherence
between basis states.

We choose a fusion-tree basis because its charge labels correspond to outcomes of compatible
measurements of the total charges of groups of anyons. These measurements involve anyons
in both subsystems. For models with fusion multiplicities, charge measurements alone do not
distinguish all basis states; additional measurements within the multiplicity spaces are required.
We assume access to these measurements when defining coherence in a complete fusion-tree basis.
No such additional measurements are needed for Fibonacci anyons.

Let $\B=\{\ket{k}\}$ be a complete fusion-tree basis, as introduced
in Sec.~\ref{subsec:fibonacci-basis}, and let $Q_k=\ket{k}\bra{k}$. Dephasing in this basis gives
\begin{equation}
   \Delta^{\B}(\tilde\rho) = \sum_k Q_k \tilde\rho Q_k.
   \label{eq:global-dephasing}
\end{equation}
It preserves every basis population and removes all off-diagonal entries. The resulting state
is the maximum-entropy state compatible with those populations. Equivalently, it minimizes
relative entropy over states diagonal in $\B$. The fusion-basis coherence is therefore
\begin{align}
   C_{\rm glob}^{\B}(\tilde\rho) = \tS[\Delta^{\B}(\tilde\rho)] - \tS(\tilde\rho)
      = \tS[\tilde\rho \Vert \Delta^{\B}(\tilde\rho)].
   \label{eq:Cglobal}
\end{align}
This definition extends ordinary relative-entropy coherence to a fusion-tree basis.
Measurements in this basis can reveal how the charges of $A$ and $B$ fuse. Such information
is generally unavailable from charge-preserving local measurements, even when their
joint statistics are recorded. We next define coherence from the local measurement data
specified in Sec.~\ref{sec:anyonic-algebras}.

We now consider measurements in fixed local bases, described by the algebra $\A_{\rm ldiag}$
defined in Sec.~\ref{subsec:product}. These measurements determine the joint basis populations,
but leave three types of information inaccessible: coherence between different local charges,
fusion information that local measurements cannot resolve, and coherence within each local-charge
block. We define the total bipartite coherence as the information lost under this measurement
restriction:
\begin{align}
   C^{\B_A\B_B}(\tilde\rho) = \tS[E_{\rm ldiag}(\tilde\rho)] - \tS(\tilde\rho)
      = \tS[\tilde\rho \Vert E_{\rm ldiag}(\tilde\rho)].
   \label{eq:total-C}
\end{align}
It quantifies the information lost when the state is described only by these joint local
populations. For an ordinary bipartite system, the states $\ket{i}_A \otimes \ket{j}_B$ form
a complete basis of the joint Hilbert space, so this definition gives the usual coherence
in that basis. However, for anyons, specifying the local basis states does not fully specify the
joint fusion state. The total bipartite coherence $C^{\B_A\B_B}$ therefore generally differs from
the fusion-tree-basis coherence $C_{\rm glob}^{\B}$.

Applying Eq.~\eqref{eq:staged-loss} to $\A_{\rm ldiag} \subseteq \A_{\rm tp}$ gives
\begin{equation}
   C^{\B_A\B_B}(\tilde\rho) = C_{\rm ntp}(\tilde\rho) + \Cc(\tilde\rho),
   \label{eq:two-term}
\end{equation}
where
\begin{align}
   C_{\rm ntp}(\tilde\rho) & = \tS[E_{\rm tp}(\tilde\rho)] - \tS(\tilde\rho),
   \label{eq:Cntp}\\
   \Cc(\tilde\rho) & = \tS[E_{\rm ldiag}(\tilde\rho)] - \tS[E_{\rm tp}(\tilde\rho)].
   \label{eq:Cconv}
\end{align}
Both terms are nonnegative by the reconstruction result in Sec.~\ref{sec:framework}. The first
is independent of the local basis: it measures information unavailable even to complete local
measurements. The subscript ``ntp'' refers to the non-tensor-product operator structure:
local products do not span the full physical operator space. The second is conventional
coherence of the reconstructed state. It depends on which bases the observers choose.

Changing the local measurement bases can reveal the coherence measured by $C_{\rm conv}$.
It cannot reveal $C_{\rm ntp}$, because all such measurements still belong to $\A_{\rm tp}$.
Access to that information requires observables beyond the local-product algebra.
For an ordinary bipartite system without these restrictions, $E_{\rm tp}$ is the identity
and $C_{\rm ntp}=0$.

The contribution $C_{\rm ntp}$ has a direct connection to anyonic charge entanglement (ACE).
The entropy of anyonic charge entanglement~\cite{Bonderson2017,XuZhou2023,4nd3-k6rw},
denoted by $S_{\rm ACE}$, is defined as the entropy increase under charge-line
severing: $S_{\rm ACE}(\tilde\rho) = \tS[\Omega^{A:B}(\tilde\rho)] - \tS(\tilde\rho)$.
Since $E_{\rm tp}=\Omega^{A:B}$, we obtain
\begin{equation}
   C_{\rm ntp}(\tilde\rho) = S_{\rm ACE}(\tilde\rho).
   \label{eq:ACE-identification}
\end{equation}
Thus the total bipartite coherence contains an ACE contribution, $C_{\rm ntp}$, in addition to
the conventional coherence $C_{\rm conv}$. The ACE contribution is independent of the chosen local
bases and cannot generally be identified with off-diagonal matrix elements alone. To clarify its
physical content, we use $P$ to separate it into coherence between different local charges and
the fusion information.

The first contribution is the entropy increase caused by $P$,
\begin{align}
   \Cs(\tilde\rho) = \tS[P(\tilde\rho)] - \tS(\tilde\rho) = \tS[\tilde\rho \Vert P(\tilde\rho)],
   \label{eq:Csector}
\end{align}
where the subscript ``sec'' stands for sector. It measures precisely the coherence between different
local charge pairs. It vanishes when the state is already block diagonal in $(a,b)$. The additional
information lost on reconstructing from local products is
\begin{align}
   \Cf(\tilde\rho) = \tS[E_{\rm tp}(\tilde\rho)] - \tS[P(\tilde\rho)] = \tS[P(\tilde\rho) \Vert E_{\rm tp}(\tilde\rho)],
   \label{eq:Cfusion}
\end{align}
where ``fus'' stands for fusion. This term measures fusion information that survives the local-charge
measurement but cannot be obtained from local products. It includes the fusion-outcome distribution
and its correlations with the internal states of the subsystems. It can be nonzero even for a density
matrix that is diagonal in a complete fusion basis. We therefore call it fusion information:
it measures more than off-diagonal coherence.

We summarize these results in the following theorem.
\begin{theorem}[Decomposition of anyonic coherence]
\label{thm:coherence-decomposition}
For any physical anyonic state $\tilde\rho$, choose orthonormal local bases $\B_A$ and $\B_B$.
The charge contributions satisfy
\begin{equation}
   C_{\rm ntp} = \Cs + \Cf = S_{\rm ACE},
   \label{eq:ACE-split}
\end{equation}
and the total bipartite coherence decomposes as
\begin{equation}
   C^{\B_A\B_B} = \Cs + \Cf + \Cc.
   \label{eq:three-term}
\end{equation}
All three contributions are nonnegative. The quantities $C_{\rm sec}$ and
$C_{\rm fus}$ are independent of the choice of bases within the local-charge
sectors, whereas $C_{\rm conv}$ generally depends on this choice.
\end{theorem}

\subsection{Abelian anyons and group superselection rules}
\label{subsec:abelian}

For Abelian anyons, specifying $a$ and $b$ fixes their fusion outcome, and the fusion space is
one-dimensional. Once coherence between local charges has been removed, no additional fusion
information remains. Thus
\begin{equation}
   E_{\rm tp} = P, \qquad \Cf = 0, \qquad C_{\rm ntp} = \Cs.
   \label{eq:abelian}
\end{equation}
This holds for both pure and mixed states. Conventional coherence can still be present if there
are internal local degrees of freedom.

Group superselection rules provide another application of the reconstruction framework.
Consider a finite-dimensional bipartite system whose state is globally invariant under
a compact symmetry group $G$~\cite{BartlettWiseman2003,RevModPhys.79.555,Vaccaro2008,Gour2009}.
Charges are labeled by irreducible representations, and fusion corresponds to decomposing
tensor products of these representations. Products of locally invariant observables form
the accessible algebra. Averaging independently over the group transformations in each
region is the trace-preserving conditional expectation onto this algebra.
The projection $P$ removes coherence between different local representation labels;
for non-Abelian $G$, the subsequent local group average also removes information
within the representation spaces. The same hierarchy and entropy decomposition apply,
with representation dimensions playing the role of anyonic quantum dimensions.

Anyonic systems allow more general fusion rules and quantum dimensions. For example,
the Fibonacci rule gives $d_\tau^2=1+d_\tau$, with $d_\tau=(1+\sqrt{5})/2$, whereas ordinary group
representations have integer dimensions. Our framework uses the charge sectors, fusion rules,
and accessible observables, and does not require braiding operations.

\subsection{Fusion information in a diagonal Fibonacci state}
\label{subsec:two-tau}

Return to one Fibonacci anyon in each region. Both local charges are $\tau$, so there are no
distinct local-charge sectors to superpose and $P=\id$. Nevertheless, the pair can fuse
to $1$ or $\tau$. Consider
\begin{equation}
   \tilde\rho(p) = p \ket{\tau, \tau;1}\bra{\tau,\tau;1} \oplus \frac{1-p}{d_\tau}
      \ket{\tau, \tau;\tau}\bra{\tau, \tau;\tau},
   \label{eq:tau-state}
\end{equation}
where $p$ is the probability of vacuum total charge. The factor $1/d_\tau$ ensures quantum-trace
normalization. These states are all diagonal, so $C_{\rm glob}^{\B} = 0$. There are no internal
local fusion degrees of freedom, and hence $\Cs = \Cc = 0$ as well.

A joint measurement of the pair can determine $p$. Separate local measurements give the same data
for every $p$, namely a definite $\tau$ on each side. Their maximum-entropy reconstruction is
\begin{equation}
   E_{\rm tp}[\tilde\rho(p)] = \frac{I}{d_\tau^2},
   \label{eq:tau-reference}
\end{equation}
with $p_{\tau\tau}(1) = d_\tau^{-2}$ and $p_{\tau\tau}(\tau) = d_\tau^{-1}$.
The missing information is
\begin{align}
   \Cf[\tilde\rho(p)] & = D_{\rm KL}[(p,1-p) \Vert (d_\tau^{-2},d_\tau^{-1})] \nonumber\\
   & = - H_2(p) + (1+p) \log_2 d_\tau,
   \label{eq:tau-fusion}
\end{align}
where $D_{\rm KL}$ is the classical relative entropy. The fusion contribution vanishes only
when the actual distribution equals the reconstructed distribution. A pair fusing to vacuum
has $\Cf = 2\log_2 d_\tau$, despite having no off-diagonal entries.

This example explains why a finer description is needed. A global basis measurement determines
the fusion outcome; local measurements do not. The difference persists even after every off-diagonal
entry has vanished.

\section{Coherence and entanglement}
\label{sec:entanglement}

In ordinary bipartite systems, coherence has been separated into local coherence and correlated
coherence~\cite{Tan2016,Guo2017,Kraft2018}. We apply this separation to the locally reconstructed
state $E_{\rm tp}(\tilde\rho)$ and divide $C_{\rm conv}$ into the same two contributions.
We then examine how correlated coherence is related to entanglement in the anyonic system,
complementing the relation $C_{\rm ntp} = S_{\rm ACE}$ established above.

\subsection{Local coherence and correlated coherence}
\label{subsec:cc}

To obtain the correlated part of $C_{\rm conv}$, we subtract the local coherence of the two
subsystems. Let $\tilde\rho_{\rm tp} = E_{\rm tp}(\tilde\rho)$. Its marginals equal those
of $\tilde\rho$, since the reconstruction preserves all local expectation values. With $\Delta_A$
and $\Delta_B$ denoting dephasing in the selected local bases, define
\begin{align}
   \Cl(\tilde\rho) & = \tS[\Delta_A(\tilde\rho_A)] - \tS(\tilde\rho_A)\nonumber\\
   & \quad + \tS[\Delta_B(\tilde\rho_B)] - \tS(\tilde\rho_B),
   \label{eq:local-coherence}\\
   \Ccc(\tilde\rho) & = \Cc(\tilde\rho) - \Cl(\tilde\rho).
 \label{eq:correlated-coherence}
\end{align}
The second quantity is correlated coherence. In terms of the anyonic mutual
information $\widetilde{I}(\sigma)=\tS(\sigma_A)+\tS(\sigma_B) -\tS(\sigma_{AB})$~\cite{XuZhou2021},
it is
\begin{equation}
   \Ccc(\tilde\rho) = \widetilde{I}(\tilde\rho_{\rm tp}) - \widetilde{I}[E_{\rm ldiag}(\tilde\rho)].
   \label{eq:cc-MI}
\end{equation}
Thus it measures how much correlation in the local-product description is lost when only the chosen
local populations are retained. It is nonnegative: within this description, the problem is an ordinary
bipartite state with locally distinguishable charge labels, and local dephasing cannot increase
mutual information. The explicit reduction is given in Appendix~\ref{app:expectation}.

Substituting $C_{\rm conv} = C_{\rm loc} + C_{\rm cc}$ into
Theorem~\ref{thm:coherence-decomposition} gives
\begin{equation}
   C^{\B_A\B_B} = \Cl + \Ccc + \Cs + \Cf.
   \label{eq:four-term}
\end{equation}
The first two terms describe coherence within the locally reconstructed state. The last two describe
information that cannot be determined by changing local measurement bases.

\subsection{Optimized coherence and pure-state entanglement}
\label{subsec:optimization}

For an ordinary bipartite pure state, the relative-entropy coherence in the Schmidt bases equals
its entanglement entropy~\cite{TanJeong2018}. The local coherence vanishes in these bases,
so the correlated coherence also equals $S(\rho_A)$.

We now examine this relation in anyonic systems, where the total coherence also contains
charge-sector and fusion contributions. The optimization is over all orthonormal local bases
within each fixed charge sector:
\begin{equation}
   C_{\rm cc}^{\rm opt} = \min_{\B_A,\B_B}\Ccc,\qquad
   C^{\rm opt} = \min_{\B_A,\B_B}C^{\B_A\B_B}.
   \label{eq:optimized-coherences}
\end{equation}
Equivalently, we allow all charge-preserving local unitaries, not just changes of fusion tree.
By Eq.~\eqref{eq:cc-MI}, minimizing correlated coherence maximizes the classical mutual information
between the two local measurement outcomes.

The conventional and total relative-entropy entanglement are~\cite{4nd3-k6rw,bkn5-drgw}
\begin{align}
   E_{\rm CE}(\tilde\rho) & = \min_{\sigma\in\mathrm{SEP}}
      \tS[E_{\rm tp}(\tilde\rho) \Vert \sigma], \nonumber\\
   E_{\rm RE}(\tilde\rho) & = \min_{\sigma\in\mathrm{SEP}} \tS(\tilde\rho \Vert \sigma),
   \label{eq:entanglement-def}
\end{align}
where $\mathrm{SEP}$ consists of states in $\A_{\rm tp}$ with separable internal states
in each local-charge block. Thus $E_{\rm CE}$ measures entanglement in the locally reconstructed
state, while $E_{\rm RE}$ compares the original state with the same separable states.
Minimizing Eq.~\eqref{eq:app-projection-proof} over $\sigma\in\mathrm{SEP}$ and using
Theorem~\ref{thm:coherence-decomposition}, we obtain
\begin{equation}
   E_{\rm RE} = E_{\rm CE}+C_{\rm ntp} = E_{\rm CE} + \Cs + \Cf.
 \label{eq:entanglement-split}
\end{equation}
The first equality is the entanglement decomposition of Ref.~\cite{4nd3-k6rw}.
The second follows from Theorem~\ref{thm:coherence-decomposition}.

For pure states with vacuum total charge, the optimized coherence
recovers these entanglement measures as follows.
\begin{theorem}[Optimized coherence and pure-state entanglement]
\label{thm:pure-coherence-entanglement}
Consider a pure anyonic state with vacuum total charge,
\begin{equation}
   \ket\Psi = \sum_a \sqrt{p_a} \ket{\psi_a}_{AB} \otimes \ket{a,\bar a;1},
      \qquad \tilde\rho = \ket\Psi\bra\Psi,
   \label{eq:vacuum-state}
\end{equation}
where $\bar a$ is the dual charge, $\sum_a p_a = 1$, and $\ket{\psi_a}$ is a normalized state
in $\mathcal{H}_A^a \otimes \mathcal{H}_B^{\bar a}$. Let $\rho_A^a = \Tr_B\ket{\psi_a}\bra{\psi_a}$
and $H(p) = -\sum_a p_a \log_2 p_a$. The charge contributions are
\begin{equation}
   \Cs(\tilde\rho) = H(p),\qquad \Cf(\tilde\rho) = 2\sum_a p_a \log_2 d_a.
   \label{eq:pure-charge-split}
\end{equation}
Optimization over local bases within each charge sector gives
\begin{equation}
   C_{\rm cc}^{\rm opt}(\tilde\rho) = \sum_a p_a S(\rho_A^a) = E_{\rm CE}(\tilde\rho),
 \label{eq:Ccc-ECE}
\end{equation}
and
\begin{align}
   C^{\rm opt}(\tilde\rho) & = H(p) + 2\sum_a p_a \log_2 d_a + \sum_a p_a S(\rho_A^a) \nonumber\\
   & = E_{\rm RE}(\tilde\rho).
   \label{eq:Copt-ERE}
\end{align}
Both minima are attained by choosing Schmidt bases for $\ket{\psi_a}$.
These bases also give $C_{\rm loc}=0$.
\end{theorem}

The proof is given in Appendix~\ref{app:pure-proof}. Reconstruction preserves the internal
pure state $\ket{\psi_a}$ in each charge sector. Measuring the local charges and then
using Schmidt bases retains the classical correlation between $a$ and $\bar a$ and
maximizes the mutual information of the internal measurement outcomes.
The remaining mutual-information loss is the average internal entanglement.

For comparison, the asymptotic entanglement entropy introduced in Ref.~\cite{Kato2014} gives
\begin{align}
   E_{\rm AEE}(\tilde\rho) & = \tS(\tilde\rho_A) \nonumber\\
   & = H(p) + \sum_a p_a[S(\rho_A^a) + \log_2d_a],
   \label{eq:pure-marginal}
\end{align}
for the same vacuum pure state.
The optimized total coherence $C^{\rm opt}=E_{\rm RE}$ therefore exceeds
$E_{\rm AEE}$ by $\sum_a p_a\log_2d_a$. This difference is consistent
with the asymptotic meaning of $E_{\rm AEE}$: for a finite number of copies,
$E_{\rm RE}$ per copy need not equal $E_{\rm AEE}$, but it converges to
$E_{\rm AEE}$ in the asymptotic limit~\cite{bkn5-drgw}.

\section{Summary and discussion}
\label{sec:conclusion}

We have studied anyonic coherence through the information available to restricted measurements.
When the accessible observables form a subalgebra, the trace-preserving conditional expectation
gives both the maximum-entropy state consistent with the data and the closest state in that
subalgebra in relative entropy. The construction also applies to other systems whose
measurement restrictions are described by observable subalgebras. The physical meaning
of each contribution is determined by which observables are retained at each step.

For a bipartite anyonic system, we distinguish three restrictions: preserving local charges,
retaining only local-product observables, and measuring in fixed local bases. The corresponding
subalgebras give a decomposition of total coherence into local-charge-sector coherence,
fusion information, and conventional coherence. In particular, the entropy of anyonic charge
entanglement is $S_{\rm ACE} = C_{\rm ntp} = C_{\rm sec} + C_{\rm fus}$. The projection $P$
removes coherence between different local charges while preserving fusion information. The latter
can be nonzero for states diagonal in a fusion-tree basis and vanishes for Abelian anyons.
We further separate conventional coherence into local and correlated contributions. For vacuum pure
states, optimizing the local bases gives the conventional relative-entropy entanglement from
correlated coherence and the total relative-entropy entanglement from total coherence.

These results also suggest a connection to topological entanglement
entropy (TEE)~\cite{KitaevPreskill2006,LevinWen2006}. Our main analysis concerns the effective
anyonic degrees of freedom, whereas TEE characterizes entanglement in a microscopic ground state.
Anyonic charge entanglement provides a bridge between these descriptions: Ref.~\cite{Bonderson2017}
derives TEE using the anyonic correlations across a spatial boundary and the constraint on their
total charge. The anyons along the boundary must have total vacuum charge, so their charges are not
independent. This constraint reduces the boundary entropy.
Building on this connection, our framework offers two observations.

First, the decomposition of $C_{\rm ntp}$ distinguishes the contribution associated with local-charge
superselection from the additional information associated with non-Abelian fusion.
For Abelian anyons, all quantum dimensions are trivial $d_a = 1$ and $C_{\rm fus} = 0$,
but local charge constraints remain. Thus non-Abelian fusion is not required for nonzero TEE:
in the toric code~\cite{Kitaev2003}, all anyons have unit quantum dimension,
yet $\gamma = \log_2 \mathcal{D} = 1$.
Here TEE arises from the constraint that the boundary anyons have total vacuum charge.

Second, TEE can be written in terms of optimized coherence of the microscopic state.
Let $\ket\Phi$ be a pure ground state of a spin or qubit system, and let $C_\Phi^{\rm opt}(R)$
be its relative-entropy coherence minimized over bases $\{\ket{i}_R\otimes\ket{j}_{\bar R}\}$
of the two regions. For a pure state, the minimum coherence equals the entanglement
entropy~\cite{TanJeong2018}:
\begin{equation}
   C_\Phi^{\rm opt}(R) = S(\rho_R).
   \label{eq:regional-identity}
\end{equation}
Each regional basis may involve all degrees of freedom within that region. This identity follows
from optimization in Schmidt bases and does not require topological order. Using the annular
partition of Ref.~\cite{LevinWen2006}, the conditional mutual information can be written as
\begin{align}
   I(A:C|B)_\Phi & = C_\Phi^{\rm opt}(AB) + C_\Phi^{\rm opt}(BC) \nonumber\\
   & \quad - C_\Phi^{\rm opt}(B) - C_\Phi^{\rm opt}(ABC).
   \label{eq:TEE-identity}
\end{align}
For the standard fixed-point ground states of quantum double and string-net models
on the plane, this combination equals $2\gamma$, where $\gamma=\log_2\mathcal D$
and $\mathcal D=(\sum_a d_a^2)^{1/2}$, for regions sufficiently large compared with
the microscopic length scales~\cite{LevinWen2006,Kim2023TEE,PhysRevB.110.165154}.
For more general ground states, spurious contributions can persist even in large regions
and at zero correlation length~\cite{Williamson2019TEE,Kim2023TEE}; a choice of partition
that excludes them for a class of stabilizer codes is given in Ref.~\cite{Han2026TEE}.
Equation~\eqref{eq:TEE-identity} expresses the conditional mutual information through
independently optimized regional coherences. One half of this quantity gives the universal
TEE when spurious contributions are absent. The basis optimization does not itself remove them.

\begin{acknowledgments}
   C.-Q. Xu is supported by the National Key Research and Development Program of
   China (Grant No.~2025YFA1411600) and by NSFC (Grant No. 12405018).
   D. L. Zhou is supported by the Science Challenge Project (Grant No. TZ2025017)
   and the National Key Research and Development Program of China (Grants No. 2021YFA1402104).
\end{acknowledgments}

\appendix

\section{Reconstruction maps}
\label{app:expectation}

\subsection{Trace-preserving conditional expectation}

We first use the ordinary trace. Let $\A$ be a finite-dimensional observable subalgebra
of the allowed operator space $\M$. Orthogonal projection onto $\A$ for the
inner product $\langle X,Y\rangle = \Tr(X^\dagger Y)$ defines a linear map $E_{\A}$
satisfying Eq.~\eqref{eq:pairing}. It fixes $\A$ and preserves the trace, as follows by taking $O=I$.

For $A_1,A_2,O\in\A$, cyclicity of the trace gives
\begin{align}
   \Tr[E_{\A}(A_1 X A_2) O] & = \Tr(X A_2 O A_1)\nonumber\\
   & = \Tr[A_1 E_{\A}(X) A_2 O],
\end{align}
which implies
\begin{equation}
   E_{\A}(A_1 X A_2) = A_1 E_{\A}(X) A_2.
   \label{eq:bimodule}
\end{equation}
This is the bimodule property.

We next show that $E_{\A}$ maps positive operators to positive operators. Since $\A$ is closed
under Hermitian conjugation, the orthogonal projection
satisfies $E_{\A}(X^\dagger)=E_{\A}(X)^\dagger$. Hence, if $X\geq0$, then $E_{\A}(X)$ is Hermitian.
Suppose that it has negative eigenvalues, and let $q$ be the projector onto the corresponding
eigenspace. Because $E_{\A}(X) \in \A$, its spectral projector $q$ also belongs to $\A$.
The definition of the orthogonal projection then gives
\begin{equation}
   \Tr[E_{\A}(X) q] = \Tr(Xq).
\end{equation}
The left-hand side is negative by the definition of $q$, whereas the right-hand side is nonnegative
because both $X$ and $q$ are positive. This contradiction proves that $E_{\A}(X) \ge 0$. The same
argument after adjoining an arbitrary finite-dimensional auxiliary system proves complete positivity.

Finally, the bimodule property and trace preservation imply that $E_{\A}$ preserves
the expectation value of every observable $O \in \A$:
\begin{equation}
   \Tr[E_{\A}(X) O] = \Tr[E_{\A}(X O)] = \Tr(X O).
\end{equation}
The expectation-value condition determines the map uniquely: if two operators in $\A$
give the same expectation value for every $O \in \A$, their difference $Z$
satisfies $\Tr(Z^\dagger Z)=0$, so $Z=0$.

For anyonic states, the same construction uses the quantum trace $\tTr$ in place of $\Tr$.
The argument is unchanged. Trace preservation then means preservation of $\tTr$.

\subsection{Maximum entropy and minimum relative entropy}
\label{app:reconstruction-proof}

We prove Theorem~\ref{thm:reconstruction} using only preservation of measurement statistics
and nonnegativity of relative entropy. Write $\bar\rho = E_{\A}(\rho)$.
By construction, $\bar\rho$ is a normalized positive operator in $\A$, and
\begin{equation}
   \Tr(\rho O) = \Tr(\bar\rho O),\qquad \forall O \in \A.
   \label{eq:app-data}
\end{equation}
Thus $\bar\rho$ is an allowed candidate for both reconstruction problems:
it belongs to the accessible subalgebra and reproduces all the measured expectation values.
We now show that $\bar{\rho}$ minimizes the relative entropy and maximizes the entropy.

\paragraph{Entropy increase under reconstruction.}
First suppose that $\bar\rho$ has no zero eigenvalues. Since $\A$ is closed
under products and linear combinations, it also contains functions of
its Hermitian operators, including $\log_2 \bar\rho$.
Equation~\eqref{eq:app-data} therefore gives
\begin{equation}
   \Tr(\rho \log_2 \bar\rho) = \Tr(\bar\rho \log_2 \bar\rho).
   \label{eq:app-log-pairing}
\end{equation}
Substitution into relative entropy yields
\begin{align}
   S(\rho \Vert \bar\rho) & = \Tr(\rho \log_2 \rho) - \Tr(\rho \log_2 \bar\rho) \nonumber\\
   & = - S(\rho) - \Tr(\bar\rho \log_2 \bar\rho) \nonumber\\
   & = S(\bar\rho) - S(\rho).
   \label{eq:app-entropy-increase}
\end{align}
Thus, the reconstruction increases the entropy by $S(\rho \Vert \bar\rho)$.

\paragraph{Minimum relative entropy.}
Consider any state $\sigma \in \A$, initially with no zero eigenvalues.
Its logarithm also belongs to $\A$,
so $\Tr(\rho \log_2 \sigma) = \Tr(\bar\rho \log_2 \sigma)$.
Using this equality to compare the two relative entropies gives
\begin{align}
   S(\rho \Vert \sigma) - S(\rho \Vert \bar\rho) & = \Tr(\rho \log_2 \bar\rho) - \Tr(\rho \log_2 \sigma) \nonumber\\
   & = \Tr[\bar\rho (\log_2 \bar\rho - \log_2 \sigma)] \nonumber\\
   & = S(\bar\rho \Vert \sigma) \ge 0.
   \label{eq:app-projection-proof}
\end{align}
Consequently, no state in the accessible subalgebra is closer to $\rho$ than $\bar\rho$
in relative entropy. Equality requires $S(\bar\rho \Vert \sigma) = 0$,
which holds only for $\sigma = \bar\rho$. This proves both optimality and uniqueness.
The relative-entropy minimization and entropy-increase relation were
established in Ref.~\cite{Gao2020}.

\paragraph{Maximum entropy for fixed measurement data.}
Now let $\omega$ be any state compatible with the measured data.
Unlike $\sigma$ above, $\omega$ need not belong to $\A$:
it may contain additional coherence or correlations invisible to the observer. Compatibility means
\begin{equation}
   \Tr(\omega O) = \Tr(\rho O) = \Tr(\bar\rho O), \qquad \forall O \in \A.
\end{equation}
Taking $O = \log_2 \bar\rho$ and expanding the relative entropy, we have
\begin{align}
   S(\omega \Vert \bar\rho) & = - S(\omega) - \Tr(\omega \log_2 \bar\rho) \nonumber\\
   & = - S(\omega) - \Tr(\bar\rho \log_2 \bar\rho) \nonumber\\
   & = S(\bar\rho) - S(\omega) \ge 0.
   \label{eq:app-maxent-proof}
\end{align}
Thus $\bar\rho$ has at least as much entropy as every compatible state.
Equality holds only when $\omega = \bar\rho$, proving the unique maximum. In fact,
specifying the accessible data fixes all the information retained in $\bar\rho$;
any further information in a compatible state can only lower its uncertainty.
This establishes the maximum-entropy statement as a consequence of the same
trace identity, without a separate optimization calculation.

\paragraph{States with zero eigenvalues.}
Zero eigenvalues do not change these conclusions. Let $s$ project onto the eigenvectors
of $\bar\rho$ with nonzero eigenvalues. The projectors $s$ and $I-s$ belong to $\A$.
The probability of the complementary subspace is therefore an accessible quantity:
\begin{equation}
   \Tr[\rho (I-s)] = \Tr[\bar\rho (I-s)] = 0.
\end{equation}
A positive density matrix with zero probability in a subspace has neither population
nor coherence connecting to that subspace. Indeed,
positivity implies $|\rho_{ij}|^2 \le \rho_{ii} \rho_{jj}$ in any orthonormal basis.
Thus $\rho = s \rho s$. The same reasoning gives $\omega = s \omega s$ for every
compatible state $\omega$. We may consequently evaluate $\log_2 \bar\rho$ on the support $s$,
where all its eigenvalues are positive, and set its extension to zero outside that support.
This extension belongs to $\A$ and all the preceding traces remain valid, with $0 \log_2 0 = 0$.

For a singular candidate $\sigma \in \A$, let $q$ be its support projector.
If $\Tr[\rho (I-q)] > 0$, then $S(\rho \Vert \sigma) = + \infty$, so it cannot improve on the finite
value at $\bar\rho$. Otherwise, preservation of this expectation value gives $\Tr[\bar\rho(I-q)]=0$
as well. Both $\rho$ and $\bar\rho$ are then supported inside $q$, and the calculation in
Eq.~\eqref{eq:app-projection-proof} applies with $\log_2\sigma$ evaluated on $q$.
This covers every candidate, including singular states, and completes the proof.

\paragraph{Anyonic trace.}
For anyonic states, replace $\Tr$ and $S$ by $\tTr$ and $\tS$, with states normalized
by $\tTr \tilde{\rho} = 1$. The proof uses cyclicity, preservation of expectation values,
and positivity of the trace, all of which hold for the quantum trace defined in the main text.
Relative entropy is also nonnegative and vanishes only for identical states. Explicitly,
for block states $\tilde{\rho} = \bigoplus_c \tilde{\rho}_c$
and $\tilde{\sigma} = \bigoplus_c \tilde{\sigma}_c$, we can define the
matrices $\rho = \bigoplus_c d_c \tilde{\rho}_c$ and $\sigma = \bigoplus_c d_c \tilde{\sigma}_c$
with ordinary trace one, and
\begin{equation}
   \tS(\tilde{\rho} \Vert \tilde{\sigma}) = S(\rho \Vert \sigma).
\end{equation}
The factors $\log_2 d_c$ cancel in the difference of logarithms. Since every $d_c$ is positive,
zero quantum-trace weight in a subspace also means zero population there.
The support arguments therefore apply unchanged. Both reconstruction statements hold.

\subsection{The local-charge projection}

The projectors $\Pi_{ab}$ describe mutually exclusive local-charge sectors
and satisfy $\sum_{a,b}\Pi_{ab}=I$. Inserting this identity on both sides of an operator gives
\begin{equation}
   X = \sum_{a,b} \sum_{a',b'} \Pi_{ab} X \Pi_{a'b'}.
\end{equation}
The map $P$ retains the terms with $(a,b)=(a',b')$ and removes the terms connecting
different local-charge sectors. Therefore, $P(X) = X$ exactly when $X$ is block diagonal
in $(a,b)$. This proves that the range of $P$ is $\A_{\rm bd}$ and that $P^2=P$.

The map $P$ is the state transformation produced by measuring the local charges and discarding
the outcome. Each term $\Pi_{ab} X \Pi_{ab}$ has Kraus form, so $P$ is completely positive.
The completeness of the projectors also gives $P(I)=I$. Moreover, cyclicity of the quantum trace
within each global-charge sector gives
\begin{align}
   \tTr[P(X)] = \sum_{a,b} \tTr(X \Pi_{ab}) =\tTr X.
\end{align}
Thus $P$ preserves the normalization of anyonic states.

Finally, every $A \in \A_{\rm bd}$ is block diagonal and hence commutes with all $\Pi_{ab}$.
For $A_1, A_2 \in \A_{\rm bd}$,
\begin{align}
   P(A_1 X A_2) & = \sum_{a,b} \Pi_{ab} A_1 X A_2 \Pi_{ab} \nonumber\\
   & = A_1 \left(\sum_{a,b} \Pi_{ab} X \Pi_{ab}\right) A_2 = A_1 P(X )A_2.
\end{align}
Thus $P$ fixes $\A_{\rm bd}$, preserves the quantum trace, and satisfies the bimodule property.
It is therefore the quantum-trace-preserving conditional expectation onto $\A_{\rm bd}$.

To make the information loss explicit, choose bases $\alpha,\beta$ of the local internal spaces
and $\mu$ of the global fusion space. On a matrix unit with fixed global charge,
\begin{align}
   & P \big(\ket{a,b; \alpha,\beta; c,\mu} \bra{a',b'; \alpha',\beta'; c,\nu}\big) \nonumber\\
   & \quad = \delta_{aa'} \delta_{bb'} \ket{a,b;\alpha,\beta;c,\mu} \bra{a',b';\alpha',\beta';c,\nu}.
   \label{eq:P-units}
\end{align}

\subsection{Charge-line severing as a conditional expectation}
\label{app:local-products}

We prove that the charge-line severing map shown diagrammatically in
Sec.~\ref{subsec:product} has the conditional-expectation property stated in
Theorem~\ref{thm:severing}. The coefficient in that graphical expression
uses the normalization of Ref.~\cite{Bonderson2017}.

\begin{proof}[Proof of Theorem~\ref{thm:severing}]
After severing, each diagram is a product of an operator in $A$ and an operator in $B$.
Thus $\Omega^{A:B}(X) \in \A_{\rm tp}$. Conversely, local-product diagrams have no nontrivial
charge line between the two subsystems, so severing leaves every $O \in \A_{\rm tp}$ unchanged.

We next check the measurement data. The quantum trace closes an operator diagram,
and $\Omega^{A:B}$ is Hermitian. In the closed diagram for $\tTr[\Omega^{A:B}(X)O]$,
the same projector can therefore be placed on the $O$ side of the contraction. This gives
\begin{align}
   \tTr[\Omega^{A:B}(X) O] & = \tTr[X \Omega^{A:B}(O)] \nonumber\\
   & = \tTr(X O),\qquad \forall O \in \A_{\rm tp}.
   \label{eq:Etp-pairing}
\end{align}
Taking $O = I$ also proves quantum-trace preservation. Hence $\Omega^{A:B}(X)$ belongs
to $\A_{\rm tp}$ and has the required expectation values for every operator in that algebra.
The uniqueness proved in this appendix implies $\Omega^{A:B} = E_{\rm tp}$.
\end{proof}

The theorem allows us to apply the maximum-entropy and relative-entropy results directly to
charge-line severing. It also clarifies the difference from $P$: the projection $P$ removes
coherence between local charges, while $\Omega^{A:B}$ removes all nontrivial charge
lines between the regions.

Dephasing the internal states of $E_{\rm tp}(\tilde\rho)$ in the chosen local bases then
gives $E_{\rm ldiag}(\tilde\rho)$. Since the local diagonal algebra is contained
in $\A_{\rm tp}$, $E_{\rm ldiag} E_{\rm tp} = E_{\rm ldiag}$ by Eq.~\eqref{eq:tower}.
Both $P$ and $E_{\rm tp}$ satisfy $E(UXU^\dagger) = UE(X)U^\dagger$ for charge-preserving
local unitaries $U$. This follows from their bimodule property, since $U$ belongs to
both algebras. Unitary invariance of relative entropy therefore proves
that $C_{\rm sec}$ and $C_{\rm fus}$ are unchanged by these unitaries.

For completeness, the local-product data can be represented by an ordinary density
matrix $\sigma = \bigoplus_{a,b} p_{ab} \rho_{ab}$, where $p_{ab}$ is the local-charge
probability and $\rho_{ab}$ is the normalized internal state determined by those data.
This representation preserves every local-product expectation value. The anyonic entropy
of the reconstructed state is
\begin{equation}
   \tS[E_{\rm tp}(\tilde\rho)] = S(\sigma)+\sum_{a,b}p_{ab}\log_2(d_a d_b).
\end{equation}
The marginal entropies have the corresponding terms $\sum_a p_a\log_2d_a$
and $\sum_b p_b\log_2d_b$, with $p_a=\sum_b p_{ab}$ and $p_b=\sum_a p_{ab}$. These terms cancel
in mutual information, giving $\widetilde I[E_{\rm tp}(\tilde\rho)] = I(\sigma)$.
Local dephasing cannot increase $I(\sigma)$, which proves the nonnegativity of correlated
coherence in Eq.~\eqref{eq:cc-MI}.




\section{Ordinary coherence and relative-entropy discord}
\label{app:ordinary}

With the ordinary trace, the same framework applies to conventional quantum systems.
For an orthonormal basis $\B=\{\ket i\}$, the diagonal algebra
is $\A_{\rm diag} = {\rm span}\{\ket i\bra i\}$, with conditional expectation
\begin{equation}
   \Delta_{\B}(\rho) = \sum_i \ket i\bra i \rho \ket i\bra i.
\end{equation}
Equation~\eqref{eq:entropy-increase} gives the relative-entropy coherence~\cite{Baumgratz2014},
\begin{equation}
   C_r^{\B}(\rho) = \min_{\delta \in \mathcal{S}(\A_{\rm diag})}
      S(\rho \Vert \delta) = S[\Delta_{\B}(\rho)] - S(\rho).
\end{equation}

For a conventional bipartite system, choosing a product basis gives
the conditional expectation $\Delta_{\B_A} \otimes \Delta_{\B_B}$ onto its local diagonal algebra.
Optimizing over both bases gives the two-sided relative-entropy discord~\cite{Yao2015},
\begin{align}
   D_{\rm RE}^{AB}(\rho) & = \min_{\chi \in \mathrm{CC}} S(\rho \Vert \chi) \nonumber\\
   & = \min_{\B_A,\B_B} \{S[(\Delta_{\B_A} \otimes \Delta_{\B_B})\rho] - S(\rho)\},
   \label{eq:RE-discord}
\end{align}
where $\mathrm{CC}$ denotes the set of classical states defined in Ref.~\cite{Modi2010},
namely, states diagonal in a local product basis. If only $A$ is dephased, the relevant algebra is
\begin{equation}
   \A_{\B_A}^{\rm block} = \bigoplus_i \ket i\bra i \otimes \mathcal{B}(\mathcal{H}_B),
\end{equation}
which gives the one-sided relative-entropy discord.

The relative-entropy discord in Eq.~\eqref{eq:RE-discord} should be distinguished
from the optimized mutual-information loss used in Sec.~\ref{subsec:optimization}.
For a fixed basis, the latter subtracts the local coherence.
These two optimizations agree on bipartite pure states but need not agree on mixed states.

\section{Correlated coherence and conventional entanglement}
\label{app:pure-proof}

In this appendix we prove that, for a pure anyonic state with vacuum total charge,
the optimized correlated coherence equals the conventional entanglement contribution.
Following Eq.~\eqref{eq:cc-MI}, we calculate
\begin{equation}
   \Ccc(\tilde\rho) = \widetilde{I}(\tilde\rho_{\rm tp}) - \widetilde{I}[E_{\rm ldiag}(\tilde\rho_{\rm tp})],
   \label{eq:app-cc-start}
\end{equation}
where $\tilde\rho_{\rm tp} = E_{\rm tp}(\tilde\rho)$. The first term is independent of
the local bases. The optimization therefore amounts to finding the largest mutual information
that remains after local dephasing.

Consider the vacuum pure state in Eq.~\eqref{eq:vacuum-state},
\begin{equation}
   \tilde\rho = \ket\Psi \bra\Psi,\qquad
   \ket\Psi = \sum_a \sqrt{p_a} \ket{\psi_a} \otimes \ket{a,\bar a;1}.
\end{equation}
Here $a$ and $\bar a$ are the local charges, $\ket{a,\bar a;1}$ describes their fusion to vacuum,
and $\ket{\psi_a}$ is a normalized ordinary bipartite pure state
in $\mathcal{H}_A^a \otimes \mathcal{H}_B^{\bar a}$. Removing coherence between local charges gives
\begin{equation}
   P(\tilde\rho) = \bigoplus_a p_a \ket{\psi_a}\bra{\psi_a} \otimes \ket{a,\bar a;1}\bra{a,\bar a;1}.
   \label{eq:vacuum-P}
\end{equation}
Each block is pure and has vacuum total charge, so $\tS[P(\tilde\rho)] = H(p)$,
where $H(p) = - \sum_a p_a\log_2p_a$.

By Theorem~\ref{thm:severing}, we have
\begin{equation}
   \tilde\rho_{\rm tp} = \Omega^{A:B}(\tilde{\rho})
      = \bigoplus_a p_a \ket{\psi_a}\bra{\psi_a} \otimes \frac{I_{a\bar a}}{d_a^2}.
 \label{eq:app-vacuum-tp}
\end{equation}
Here $I_{a\bar a}$ is the identity on $\bigoplus_c V_{a\bar a}^{c}$, including all allowed
fusion outcomes $c$. Its quantum trace is $\sum_c N_{a\bar a}^{c}d_c=d_a^2$,
so $I_{a\bar a}/d_a^2$ has unit quantum trace. The reconstruction preserves the internal
pure state in each local-charge sector but replaces the known vacuum fusion outcome
by this normalized identity.

The nonzero eigenvalue in every allowed fusion channel of sector $a$ is $p_a/d_a^2$.
Taking the quantum trace gives
\begin{align}
   \tS(\tilde\rho_{\rm tp}) & = - \sum_{a,c} N_{a\bar a}^{c} d_c \frac{p_a}{d_a^2}
      \log_2 \frac{p_a}{d_a^2} \nonumber\\
   & = H(p) + 2\sum_a p_a\log_2d_a.
   \label{eq:app-tp-entropy}
\end{align}
Together with $\tS(\tilde\rho) = 0$ and $\tS[P(\tilde\rho)] = H(p)$, this proves
the charge-sector and fusion contributions in Eq.~\eqref{eq:pure-charge-split}.

Define the internal reduced states by
\begin{equation}
   \rho_A^a = \Tr_B\ket{\psi_a}\bra{\psi_a},\qquad
   \rho_B^{\bar a} = \Tr_A\ket{\psi_a}\bra{\psi_a}.
\end{equation}
They have the same nonzero eigenvalues because $\ket{\psi_a}$ is pure.
The anyonic reduced states are
\begin{equation}
   \tilde\rho_A = \bigoplus_a\frac{p_a}{d_a}\rho_A^a,\qquad
   \tilde\rho_B = \bigoplus_a\frac{p_a}{d_a}\rho_B^{\bar a}.
\end{equation}
Taking the quantum trace in each local-charge sector gives
\begin{align}
   \tS(\tilde\rho_A) = \tS(\tilde\rho_B) = H(p) + \sum_a p_a\big[S(\rho_A^a) + \log_2d_a\big].
\end{align}
This proves Eq.~\eqref{eq:pure-marginal}. Subtracting Eq.~\eqref{eq:app-tp-entropy}
from the sum of the two reduced entropies, we obtain
\begin{equation}
   \widetilde{I}(\tilde\rho_{\rm tp}) = H(p) + 2\sum_a p_aS(\rho_A^a).
   \label{eq:MI-pure}
\end{equation}
The quantum dimension terms cancel.

Local dephasing preserves the charge labels and dephases the internal state within
each sector. Choose bases $\{\ket{i_a}_A\}$ and $\{\ket{j_{\bar a}}_B\}$, and let
\begin{equation}
   q_{ij|a} = \left|{}_A\!\bra{i_a}{}_B\!\bra{j_{\bar a}} \psi_a\rangle\right|^2
\end{equation}
be the joint probability of their outcomes at fixed $a$.
Writing $\ket{i_a j_{\bar a}} = \ket{i_a}_A \otimes \ket{j_{\bar a}}_B$,
the dephased reconstruction is
\begin{equation}
   E_{\rm ldiag}(\tilde\rho_{\rm tp}) = \bigoplus_a p_a \sum_{i,j}q_{ij|a}
      \ket{i_a j_{\bar a}}\bra{i_a j_{\bar a}} \otimes \frac{I_{a\bar a}}{d_a^2}.
\end{equation}
Let $q_{i|a}=\sum_jq_{ij|a}$ and $q_{j|a}=\sum_iq_{ij|a}$. The classical mutual
information within sector $a$ is
\begin{equation}
   I(q_{ij|a}) = H(q_{i|a}) + H(q_{j|a}) - H(q_{ij|a}).
\end{equation}
Evaluating the three entropies in the anyonic mutual information again
cancels the quantum dimension terms, giving
\begin{equation}
   \widetilde I[E_{\rm ldiag}(\tilde\rho_{\rm tp})] = H(p) + \sum_a p_a I(q_{ij|a}).
   \label{eq:MI-dephased}
\end{equation}

For each sector, $I(q_{ij|a})$ cannot exceed $S(\rho_A^a)$. To see why,
first measure $A$ in its chosen basis. Since $\ket{\psi_a}$ is pure and
the measurement is rank one, each outcome $i$ prepares a pure conditional
state $\rho_{B|i}^a$ on $B$. Their average is $\rho_B^{\bar a}$. The Holevo quantity
of this ensemble is therefore
\begin{equation}
   \chi_a = S(\rho_B^{\bar a}) - \sum_i q_{i|a}S(\rho_{B|i}^a) = S(\rho_A^a).
\end{equation}
By the Holevo bound~\cite{nielsen2002quantum}, a measurement on $B$ can
extract at most $\chi_a$ bits of information about $i$. Consequently,
\begin{equation}
   I(q_{ij|a}) \le S(\rho_A^a).
\end{equation}

This bound is attained in Schmidt bases. In the decomposition
\begin{equation}
   \ket{\psi_a} = \sum_\alpha \sqrt{\lambda_\alpha^a} \ket{\alpha_a}_A\ket{\alpha_{\bar a}}_B,
\end{equation}
measuring both sides in these bases
gives $q_{\alpha\beta|a} = \lambda_\alpha^a\delta_{\alpha\beta}$.
Thus $I(q_{\alpha\beta|a}) = H(\lambda^a) = S(\rho_A^a)$. The bases can be chosen
independently in each local-charge sector, so all these bounds can be attained
simultaneously:
\begin{equation}
   \max_{\B_A,\B_B}\widetilde{I}[E_{\rm ldiag}(\tilde\rho_{\rm tp})] = H(p) + \sum_a p_aS(\rho_A^a).
\end{equation}
Combining this result with Eqs.~\eqref{eq:app-cc-start} and \eqref{eq:MI-pure} gives
\begin{align}
   C_{\rm cc}^{\rm opt}(\tilde\rho) & = \left[H(p) + 2\sum_a p_a S(\rho_A^a)\right] \nonumber\\
   & \quad - \left[H(p) + \sum_a p_a S(\rho_A^a)\right] \nonumber\\
   & = \sum_a p_a S(\rho_A^a).
\end{align}

We now compare this optimized correlated coherence with the entanglement of the same state.
For the vacuum pure state, $\tS(\tilde\rho)=0$, so Eq.~\eqref{eq:app-tp-entropy} gives
\begin{equation}
   C_{\rm ntp}(\tilde\rho) = H(p) + 2\sum_a p_a \log_2 d_a.
\end{equation}
For the entanglement measures defined in Eq.~\eqref{eq:entanglement-def}, the pure-state
result of Ref.~\cite{bkn5-drgw} and the decomposition in Eq.~\eqref{eq:entanglement-split} give
\begin{align}
   E_{\rm CE}(\tilde\rho) & = \sum_a p_a S(\rho_A^a), \nonumber\\
   E_{\rm RE}(\tilde\rho) & = H(p) + 2\sum_a p_a \log_2 d_a
      + \sum_a p_a S(\rho_A^a).
\end{align}
The result above therefore establishes $C_{\rm cc}^{\rm opt} = E_{\rm CE}$.

Since $C_{\rm conv} = C_{\rm loc} + C_{\rm cc}$ and $C_{\rm loc} \ge 0$,
we have $C_{\rm conv} \ge C_{\rm cc}^{\rm opt}$. The Schmidt bases attain this bound:
they minimize $C_{\rm cc}$ and also give $C_{\rm loc} = 0$.
Thus $\min_{\B_A,\B_B} C_{\rm conv} = E_{\rm CE}$. Because $C_{\rm ntp}$ is basis independent,
Theorem~\ref{thm:coherence-decomposition} yields
\begin{equation}
   C^{\rm opt}(\tilde\rho) = C_{\rm ntp}(\tilde\rho) + E_{\rm CE}(\tilde\rho) = E_{\rm RE}(\tilde\rho),
\end{equation}
which proves Eq.~\eqref{eq:Copt-ERE}.

\bibliography{AC}
\end{document}